\documentclass[twocolumn,prb,floatfix,superscriptaddress,nobibnotes, maxnames=15]{revtex4-2}
\usepackage{graphicx}
\usepackage{enumerate}
\usepackage[colorlinks=true,citecolor=blue]{hyperref}
\usepackage{textcomp}
\usepackage{amsmath}
\usepackage{amssymb}

\usepackage{pgf}
\usepackage{subfigure}
\usepackage[english]{babel}
\usepackage[normalem]{ulem}
\usepackage{tabularray}
\usepackage{xcolor}
\usepackage{array}

\graphicspath{{./}{./Figures/}}

\begin{document}

\title{Subgap transport in superconductor--semiconductor hybrid islands: Weak and strong coupling regimes}

\author{Marco Valentini$^{*}$}
\affiliation{Institute of Science and Technology Austria, Am Campus 1, 3400 Klosterneuburg, Austria.}
\author{Rubén Seoane Souto$^{*}$}
\affiliation{Instituto de Ciencia de Materiales de Madrid, Consejo Superior de Investigaciones Científicas (ICMM-CSIC), Madrid, Spain.}
\affiliation{NanoLund and Solid State Physics, Lund University, Box 118, 22100 Lund, Sweden.}
\affiliation{Center for Quantum Devices, Niels Bohr Institute, University of Copenhagen, 2100 Copenhagen, Denmark.}
\author{Maksim Borovkov}
\affiliation{Institute of Science and Technology Austria, Am Campus 1, 3400 Klosterneuburg, Austria.}
\author{Peter Krogstrup}
\affiliation{Center for Quantum Devices, Niels Bohr Institute, University of Copenhagen, 2100 Copenhagen, Denmark.}
\affiliation{NNF Quantum Computing Programme, Niels Bohr Institute, University of Copenhagen, Universitetsparken 5, 2100 Copenhagen, Denmark.}
\author{Yigal Meir}
\affiliation{Department of Physics, Ben-Gurion University of the Negev, Beer-Sheva 84105, Israel.}
\author{Martin Leijnse}
\affiliation{NanoLund and Solid State Physics, Lund University, Box 118, 22100 Lund, Sweden.}
\author{Jeroen Danon}
\affiliation{Department of Physics, Norwegian University of Science and Technology, NO-7491 Trondheim, Norway.}
\author{Georgios Katsaros}
\affiliation{Institute of Science and Technology Austria, Am Campus 1, 3400 Klosterneuburg, Austria.}

\date{\today  \ \ \  $^*$ Equal contribution. }

\begin{abstract}
Superconductor--semiconductor hybrid systems play a crucial role in realizing nanoscale quantum devices, including hybrid qubits, Majorana bound states, and Kitaev chains.
For such hybrid devices, subgap states play a prominent role in their operation. In this work, we study such subgap states via Coulomb and tunneling spectroscopy through a superconducting island defined in a semiconductor nanowire fully coated by a superconductor.
We systematically explore regimes ranging from an almost decoupled island to the open configuration.
In the weak coupling regime, the experimental observations are very similar in the absence of a magnetic field and when one flux quantum is piercing the superconducting shell.
Conversely, in the strong coupling regime, significant distinctions emerge between the two cases.
We ascribe this different behavior to the existence of subgap states at one flux quantum, which become observable only for sufficiently strong coupling to the leads. 
We support our interpretation using a simple model to describe transport through the island.
Our study highlights the importance of studying a broad range of tunnel couplings for understanding the rich physics of hybrid devices.

\end{abstract}

\maketitle

\section{Introduction}

\begin{figure*}
    \includegraphics[]{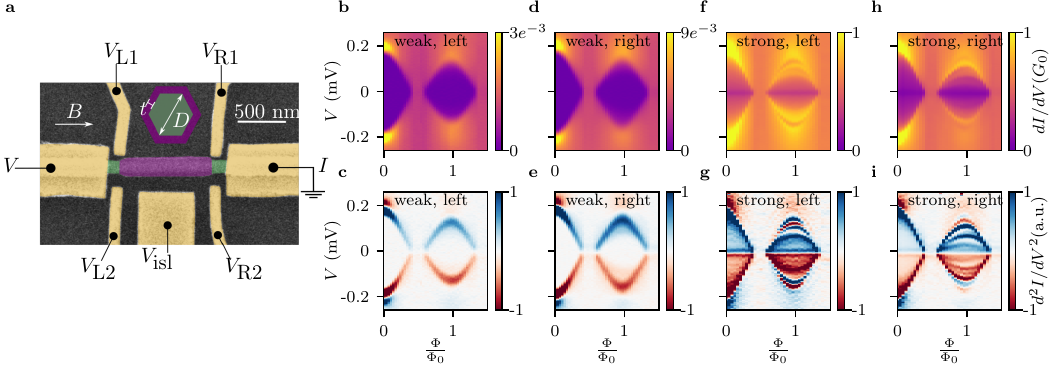}    \caption{\textbf{Device layout and tunneling spectroscopy in the weak and strong coupling regimes.} \textbf{a} False-color scanning electron micrograph of device B. Purple corresponds to Al, green to InAs and yellow to Au. The inset represents a sketch of the nanowire cross section, with $t \approx 25$ nm and $D \approx 110$ nm. The bias (current) is always applied (measured) to (from) the same lead; the tunnel barrier configuration determines the side on which the spectroscopy measurement is performed. 
    \textbf{b} [\textbf{c}] Differential conductance $dI/dV$ [second derivative $d^2I/dV^2$] of device A as a function of voltage bias $V$ and flux $\Phi$ with the left side tuned to weak coupling ($G_{\textrm{L}} \approx 0.001\, G_0$) and the right side tuned to be fully transparent (resulting in $E_{\textrm{c}} \approx 0$).  \textbf{d} [\textbf{e}] $dI/dV$ [$d^2I/dV^2$] as a function of $V$ and $\Phi$ with the right side tuned to weak coupling ($G_{\textrm{R}} \approx  0.004\, G_0$) and the left side fully transparent.  \textbf{f} [\textbf{g}] $dI/dV$ [$d^2I/dV^2$] as a function of $V$ and $\Phi$ with the left side tuned to strong coupling ($G_{\textrm{L}} \approx  0.65\, G_0$) and the right side fully transparent. \textbf{h} [\textbf{i}] $dI/dV$ [$d^2I/dV^2$] as a function of $V$ and $\Phi$ with the right side tuned to strong coupling ($G_{\textrm{R}} \approx  0.6\, G_0$) and the left side fully transparent.
    A Savitzky--Golay filter was used to improve the readability of the $d^2I/dV^2$ maps. 
    }
    \label{fig_0}
\end{figure*}

Improvements in the growth of semiconductor--superconductor heterostructures~\cite{krogstrup2015epitaxy,shabani2016two,cheah2023control}, combined with the introduction of new material combinations~\cite{sestoft2018engineering,kanne2021epitaxial,pendharkar2021parity,mazur2022spin,tosato2023hard,aggarwal2021enhancement}, novel nanofabrication techniques~\cite{carrad2020shadow,goswami2023sn,heedt2021shadow,borsoi2021single,levajac2023subgap}, and new measurement methods~\cite{van2017microwave,tosi2019spin,menard2020conductance} have led to groundbreaking hybrid devices, such as
%
Cooper pair splitters~\cite{wang2022singlet,wang2023triplet,bordoloi2022spin}, minimal Kitaev chains~\cite{dvir2023realization,tsintzis2022creating,tsintzis2023roadmap}, Andreev spin qubits~\cite{hays2021coherent,pita2023direct} and gate-tunable transmon qubits~\cite{de2015realization,larsen2015semiconductor,casparis2018superconducting,sabonis2020destructive,hertel2022gate}. These devices share a common characteristic: their working principle is based on the presence of Andreev bound states, i.e., subgap states which are a natural consequence of a non-uniform superconducting order parameter~\cite{Prada_review}. In particular, subgap states in hybrid semiconductor--superconductor systems are utilized to couple spin-polarized quantum dots to realize minimal Kitaev chains~\cite{dvir2023realization,tenHaaf2024, tsintzis2022creating,tsintzis2023roadmap,PhysRevLett.132.056602,Souto_arXiv2024} that can be used to encode quantum information~\cite{Nitsch_arXiv2024}. Extending these systems to create longer Kitaev chains~\cite{Bordin_arXiv24,ten2024edge} and demonstrating non-Abelian statistics through braiding and fusion schemes will require precise control over these subgap states. 

In addition, such subgap states are also a recurring theme in semiconducting nanowires fully wrapped by a superconducting shell (full-shell) nanowires. These nanowires exhibit a modulation of the superconducting order parameter as a function of the magnetic flux piercing the superconducting shell, known as the Little--Parks effect~\cite{LP_1962,schwiete_2010,vaitieknas2019anomalous}. Signatures of low-energy states in full-shell nanowires have been previously reported and interpretations based on Majorana zero modes or trivial states have been provided~\cite{Vaitiekenas_Science2020,razmadze2020quantum,razmadze2023supercurrent,Valentini2021,valentini2022majorana,escribano2022fluxoid,Paya_arXiv2024}. However, so far most of these studies have not investigated in detail the impact of the transparency of the tunnel barriers on the experimental manifestation of the zero- and high-bias features.

In this work, we study the transport properties of InAs/Al full-shell nanowires.
Tunneling and Coulomb blockade spectroscopy measurements were performed for different tunnel coupling configurations and magnetic field strengths; we report data from a 400 nm long (device A) and a 900 nm long proximitzed island (device B).
By comparing finite-bias spectroscopy in the different coupling regimes, we show the existence of weakly coupled subgap states at one flux quantum. We explain their impact on the experimentally observed features for a large parameter space. We find that weakly coupled subgap states block current for bias voltages below twice the gap. In this regime, the current above the gap is allowed by Cooper pair recombination. In the strong coupling regime, subgap features appear, signaling transport mediated by subgap states.

\section{Setup and tunneling spectroscopy}

A scanning electron microscopy image of a typical device is shown in Fig.~\ref{fig_0}\textbf{a}.
The superconducting island (purple) is defined by wet etching and the potential in the bare sections of InAs (green) can be tuned via nearby gates. In particular, the normal-state conductances of the left and right barriers ($G_{\textrm{L}}$ and $G_{\textrm{R}}$) are independently controlled by applying voltages $V_{\textrm{L1,2}}$ and $V_{\textrm{R1,2}}$ as indicated in the figure.
We say that a barrier is tuned to the weak coupling regime if its associated conductance $G_{\textrm{R}}$ or $G_{\textrm{L}}$ is smaller than $0.05\, G_0$ (where $G_0 = \frac{2 e^2}{h}$ is the conductance quantum), otherwise we will refer to it as strong coupling \footnote{In Ref.~\cite{valentini2022majorana} this regime was dubbed as intermediate coupling. We furthermore note that the conductances refer to the single barrier conductances and have been determined at zero flux (see supplementary information of Ref.~\cite{valentini2022majorana}).}.
The equilibrium charge occupancy of the island can be varied through the island gate voltage $V_{\textrm{isl}}$.

We first show local tunneling spectroscopy data for device A, i.e., the 400 nm long island (for device B the data are shown in App.~\ref{app:data}; see Fig.~\ref{figuresup_tun}). Measurements were carried out using standard ac lock-in techniques in a dilution refrigerator with a three-axis vector magnet and base temperature of 20 mK.
We open one side ($G_{\textrm{R}} \gg 2\, G_0$), such that the charging energy of the island $E_{\textrm{c}}=e^2/2C$ is suppressed, where $e$ is the electron charge and $C$ the total island capacitance.
We keep the other side of the island in the weak coupling regime ($G_{\textrm{L}} \approx 0.001\, G_0$) to measure the local density of states. Figure~\ref{fig_0}\textbf{b} shows the differential conductance at the left side, $dI/dV$, as a function of voltage bias $V$ and the magnetic flux $\Phi$ threading the nanowire. At zero field we see a clear superconducting gap $\Delta_0 \approx 180~\mu$eV. When increasing the field, the superconducting gap $\Delta$ is modulated due to the Little--Parks effect~\cite{LP_1962,schwiete_2010,vaitieknas2019anomalous}. The gap vanishes close to $\Phi=\frac{1}{2}\Phi_0$ ($\Phi_0$ is the magnetic flux quantum), and it reopens around $\Phi=0.6\, \Phi_0$, reaching a local maximum at $\Phi=\Phi_0$. We will refer to these two superconducting regions as the zeroth lobe (0L) and first lobe (1L). No subgap states are visible in any of the two lobes. This observation is further enlightened by the second derivative $d^2I/dV^2$, shown in Fig.~\ref{fig_0}\textbf{c}, emphasizing the absence of observable subgap conductance features. A similar result is obtained if tunneling spectroscopy is performed at the right side of the device, i.e., tuning the right junction to the weak coupling regime ($G_{\textrm{R}} \approx 0.004\, G_0$) and opening the left junction, see Figs.~\ref{fig_0}\textbf{d,e}. We note that in this case, a faint feature, more clearly visible in the $d^2I/dV^2$ map, appears close to $\Delta$. We attribute this small difference between the left and right sides to slightly different tunnel barrier strengths. 

Now, we focus on spectroscopy measurements performed in the strong coupling regime. Figure~\ref{fig_0}\textbf{f} shows the data with the left junction tuned to the strong coupling regime ($G_{\textrm{L}} \approx 0.65\,G_0$) and the right junction fully transparent. In the 0L, the conductance below $\Delta$ is increased because of Andreev processes but no subgap states appear. However, in the 1L, subgap states are now visible. This is further illuminated by the sign changes in the second derivative, see Fig.~\ref{fig_0}\textbf{g}. An analogous situation was also observed on the right side with an even more pronounced presence of subgap features, see Figs.~\ref{fig_0}\textbf{h,i}.

It is challenging to comment on the origin of these subgap states with just tunneling spectroscopy measurements, i.e., using only a local probe. 
In particular, it is difficult to understand if they are localized at the InAs junctions, like Yu--Shiba--Rusinov states can be~\cite{jellinggaard2016tuning,Valentini2021} or extend along the nanowire~\cite{razmadze2023supercurrent}, such as Caroli--de Gennes--Matricon states~\cite{caroli1964bound,san2023theory,paya2023phenomenology} ~\footnote{We will refer to all observed states with an energy below the superconducting gap as ``subgap'' states, as their exact nature depends on $E_c$, $\Delta$ and $\Gamma$, which we are varying throughout this work.}. Since Coulomb spectroscopy provides additional information about the spatial spread of the subgap states involved in transport~\cite{Souto_PRB2022}, we next study
finite-bias transport in different coupling regimes. 

\section{Coulomb spectroscopy}

\begin{figure*}
    \includegraphics[width=\textwidth]{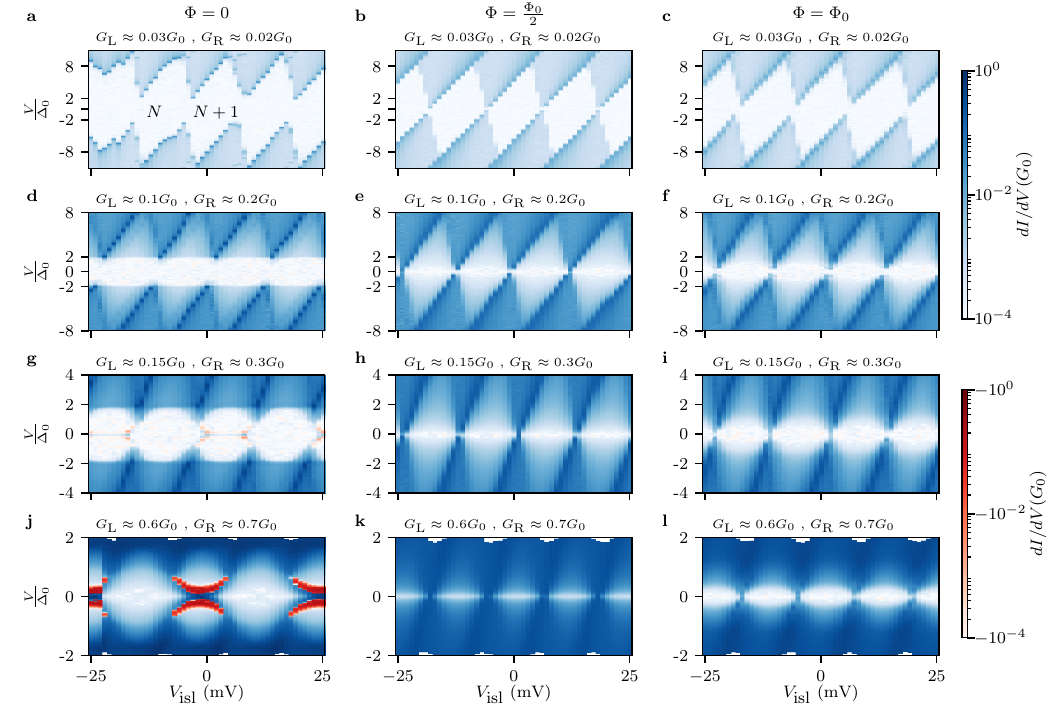}    \caption{\textbf{Coulomb spectroscopy going from the weak to strong coupling regime for different $\Phi$ and for a 900 nm long island device.}
    Each panel shows the measured differential conductance, which can be positive (blue) or negative (red), as a function of applied bias voltage $V$ and the island gate potential $V_\text{isl}$.
    The three columns of panels have $\Phi=0$, $\Phi = \frac{1}{2}\Phi_0$, and $\Phi = \Phi_0$, from left to right. The four rows have increasing $G_\text{L,R}$ from top to bottom, as indicated at the panels. The weak coupling data (top row) have $E_\text{c} \approx 4\Delta_0$ and by lowering the barriers $E_\text{c}$ gets reduced until $E_\text{c} \approx \Delta_0$ in the bottom row.
    }
    \label{fig_1}
\end{figure*}

By exploiting the full tunability of our junctions, we perform Coulomb spectroscopy in different regimes on device B, i.e., the 900 nm long island (the data for device A are shown in App.~\ref{app:data}; see Fig.~\ref{figuresup_coul}).
We start by tuning both junctions to the weak coupling regime ($G_{\textrm{L}} \approx 0.03\, G_0$ and $G_{\textrm{R}} \approx 0.02\, G_0$) and we measure the differential conductance $dI/dV$ as a function of $V$ and $V_{\textrm{isl}}$ for $\Phi=0$ (Fig.~\ref{fig_1}\textbf{a}), $\Phi=\frac{1}{2}\Phi_0$ (Fig.~\ref{fig_1}\textbf{b}) and $\Phi=\Phi_0$ (Fig.~\ref{fig_1}\textbf{c}).
In Fig.~\ref{fig_1}\textbf{b}, the Little--Parks effect destroys the superconductivity completely, meaning that the device becomes a normal single-electron transistor. These measurements allow us (i) to read off $E_\text{c} \approx 4\,\Delta_0$, where $\Delta_0 \approx 180 \ \mu$eV is the superconducting gap in the absence of a magnetic field, and (ii) to conclude that the applied bias $V$ is dropping asymmetrically over the two barriers.
Since $E_\text{c} > \Delta_0$, the ground state of the island has either an even or an odd number of electrons, depending on $V_{\textrm{isl}}$.
In this regime, charge degeneracy points exist for certain values of $V_\text{isl}$, where single electron transport is allowed at low energy, and should lead to a finite zero-bias conductance.
Surprisingly, both in the absence of a magnetic field (Fig.~\ref{fig_1}\textbf{a}) and at one flux quantum (Fig.~\ref{fig_1}\textbf{c}) no zero-bias transport signatures
are observed, despite the presence of charge-degeneracy points.
In all cases, finite conductance only appears for $e|V| > 2\Delta$, where $\Delta$ is the flux-dependent gap.
In the next section, we provide an interpretation based on the blockade of electron tunneling via the island's continuum of states.

The suppression of the zero-bias conductance persists when increasing the tunnel coupling to the leads ($G_{\textrm{L}} \approx 0.1\, G_0$ and $G_{\textrm{R}} \approx 0.2\, G_0$), second row of Fig.~\ref{fig_1}. Conductance steps across all $V_\text{isl}$ appear now at $e|V|=2\Delta$~\cite{Vaitiekenas_PRB2022,Souto_PRB2022} due to an increased contribution from inelastic cotunneling processes that leave two excited electrons above the gap and thereby open up a transport channel~\cite{Vaitiekenas_PRB2022}.
In order to see transport for $e| V | < 2 \Delta$, we need to increase the coupling further.  Fig.~\ref{fig_1}\textbf{g} shows the Coulomb diamond pattern with $G_{\textrm{L}} \approx 0.15\, G_0$ and $G_{\textrm{R}} \approx 0.3\, G_0$ for $\Phi=0$. In this case, the zero-bias transport is even--odd modulated indicating $E_{\textrm{c}} > \Delta_0$ and negative differential conductance (NDC) features appear~\cite{hekking_prl_1993,Higginbotham_NatPhys2015,Albrecht_PRL2017}. At half flux quantum (Fig.~\ref{fig_1}\textbf{h}), the island becomes normal, allowing for transport to occur at low bias. Zero-bias transport persists in the 1L (Fig.~\ref{fig_1}\textbf{i}), indicating the presence of low-energy states. These states can be either discrete or form a quasi-continuum.

If the couplings are further increased   ($G_{\textrm{L}} \approx 0.6\, G_0$ and $G_{\textrm{R}} \approx 0.7\, G_0$), such that $E_{\textrm{c}} \approx \Delta_0$, see the bottom row of Fig.~\ref{fig_1}, zero-bias transport becomes $2e$ periodic in the absence of flux, (Fig.~\ref{fig_1}\textbf{j}) \cite{Tuominen1992PRL}, while in the 1L $1e$ periodic zero-bias transport persists (Fig.~\ref{fig_1}\textbf{l}). In App.~\ref{app:data} we show additional data:
A crossover from tunneling to Coulomb spectroscopy is shown in Fig.~\ref{crossover} and
Coulomb spectroscopy for an asymmetric tuning of tunnel barriers is shown in Fig.~\ref{sup2}.
Below we proceed by developing a theoretical model to understand our Coulomb diamond data, which needs to capture our two main observations.
First, transport is suppressed for $e| V | < 2 \Delta$ in the weak coupling regime, where $E_c>\Delta$. Second, the cases $\Phi=0$ and $\Phi=\Phi_0$ give similar results in the weak coupling regime, but differ significantly in the strong coupling regime.

\begin{figure*}
    \includegraphics[width=\linewidth]{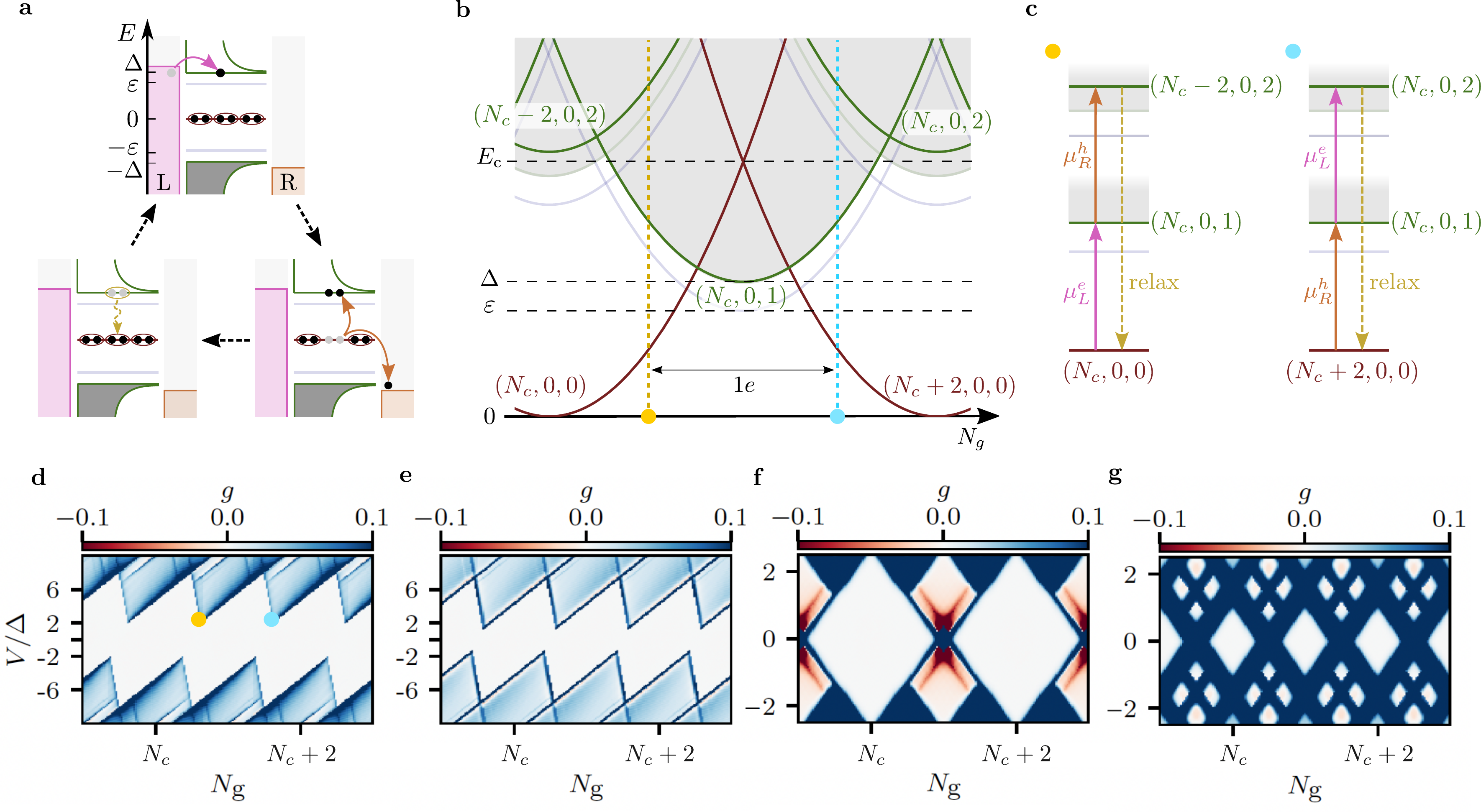}
    \caption{\textbf{Theoretical model of the hybrid island in the weak and strong coupling regimes.} \textbf{a} Transport cycle in the regime where the coupling to the subgap states is negligible. \textbf{b} Energy spectrum as a function of the island's charge offset, where we used notation $(N_c,n_s,n_c)$, with $N_c$, $n_s$, and $n_c$ being the number of electrons forming Cooper pairs and quasiparticle excitations in the subgap and the continuum of states, respectively. \textbf{c} Transport processes represented in the energy domain. \textbf{d,e} Simulated conductance in the weak coupling regime for the 0L and 1L. \textbf{f,g} Simulated conductance in the strong coupling regime, where leads couple efficiently to the island's subgap states. We consider the charging energy renormalization due to the coupling to the leads, and take $E_{\rm c}/\Delta_0=3$ and $0.8$ for the weak (\textbf{d,e}) and strong coupling (\textbf{f,g}), respectively, with $\Delta_0$ being the gap in the 0L. The gap in the 1L is taken slightly smaller and is set to $0.6\,\Delta_0$. The subgap state energy is $\varepsilon=0.9\,\Delta_0$ for the 0L and $\varepsilon = 0$ for the 1L. The tunnel rate to the continuum of states is taken as $1\cdot10^{-4}\Delta_0$ and the tunnel rate to the subgap states is negligible in the weak coupling regime and $1\cdot10^{-3}\Delta_0$ in the strong coupling regime. We used a bias asymmetry $\mu_L/\mu_R=0.15$ and $1$ for the two regimes.}. We used $k_{\rm B} T=0.025\,\Delta_0$
    \label{fig_2}
\end{figure*}

\section{Theoretical model}

To understand the qualitative transport features described above, we consider a transport model that accounts for electron pairing and electrostatic repulsion between the electrons on the island, see App.~\ref{sec:model} for details.
We allow the island to host discrete subgap states with energy $\varepsilon<\Delta$ and we use a rate-equation approach that accounts for the sequential tunneling processes between the leads and the island to describe the source-drain current.
We emphasize that we only include tunneling events to lowest order in the couplings and thus do not describe cotunneling or Andreev processes.

We begin our analysis in the weak coupling regime, where spectroscopy measurements do not show signatures of subgap states in any of the two lobes and Coulomb spectroscopy seems to suggest that transport is only possible for $e|V| \geq 2\Delta$, see Figs.~\ref{fig_0}\textbf{b}--\textbf{e} and the top row of Fig.~\ref{fig_1}.
The height of the Coulomb diamonds in Fig.~\ref{fig_1}b indicates that $E_\text{c} > \Delta_0$, as discussed before.
In this case, zero-bias conductance peaks are generally expected to appear at charge-degeneracy points where the lowest even and odd states are degenerate.
The absence of such zero-bias features suggests that the quasiparticle states that are energetically allowed to contribute to transport are not well coupled to the leads. On the other hand, the observed features at higher bias, $e|V| > 2\Delta$, suggest not only that single-electron transport is mediated by the continuum states but also that the coupling between these continuum states and the leads is in fact strong enough to induce significant tunneling (also supported by the clearly visible continuum states in Fig.~\ref{fig_0}\textbf{b,d}).

To resolve this apparent contradiction, we developed the following picture:
We assume that possible subgap states are too weakly coupled to the leads in this regime to contribute to transport.
Single electrons or holes with energies above $\Delta$ can efficiently tunnel into the above-gap continuum due to its large density of states, where the required excitation energy has to be provided by the lead. Subsequent tunneling out of the quasiparticle, however, involves only the single state it is occupying, and this process will thus be much slower~\cite{Albrecht_PRL2017}.
The excited charges in the continuum can block current flow due to electrostatic repulsion, also at the charge-degeneracy points.
Finite current can, however, arise due to the sequential excitation of two quasiparticles, after one electron tunnels from the source and a hole from the drain. The excitation processes are followed by the recombination of these quasiparticles into a Cooper pair.
This cycle of processes requires indeed a total energy of at least $2\Delta$ from the leads, and is sketched in Fig.~\ref{fig_2}\textbf{a}.
More ``conventional'' Andreev processes involving only one of the leads are suppressed by the relatively large charging energy on the island.

Using this picture, we can understand the locations of the onset of finite conductance in the $(V,N_\text{g})$-plane, where $N_{\rm g}$ represents the charge offset, proportional to $V_{\rm isl}$.
For this, we write the state of the system as $(N_c,n_s,n_c)$, where $N_c/2$ is the number of Cooper pairs, $n_c$ the number of quasiparticles in the continuum, and $n_s$ the number of quasiparticles excited in the subgap states.
The latter will only become important in the strong coupling regime.
The lowest bias voltage for which a cycle like the one sketched in Fig.~\ref{fig_2}\textbf{a} can take place is when the electrochemical potential of the electrons in the source lead $\mu_L^e$ equals the energy difference between the states marked $(N_c,0,1)$ and $(N_c,0,0)$ in Fig.~\ref{fig_2}\textbf{b,c}  and the chemical potential in the drain lead $\mu_R^h$ equals the energy difference between $(N_c-2,0,2)$ and $(N_c,0,1)$.
The energy differences of the two sequential tunneling processes are sketched in Fig.~\ref{fig_2}\textbf{c}, for the case of $N_g$ being tuned to the yellow spot in Fig.~\ref{fig_2}\textbf{b} and assuming symmetric bias, i.e., $\mu_L^e = \mu_R^h = eV/2$, for simplicity. 
The explicit conditions mentioned above
\begin{gather*}
    E_\text{c}(N_c+1-N_{\rm g})^2 + \Delta - E_\text{c} (N_c-N_{\rm g})^2 = \mu_L^e, \\
    E_\text{c} (N_c-N_{\rm g})^2 + 2\Delta - E_\text{c}(N_c+1-N_{\rm g})^2 - \Delta = \mu_R^h,
\end{gather*}
can be straightforwardly rewritten as
\begin{gather*}
    \mu_L^e + \mu_R^h = 2\Delta,\\
    N_{\rm g} = \frac{1}{2} \left( 1 + 2N_c + \frac{\Delta}{E_\text{c}} \frac{\mu_R^h - \mu_L^e}{\mu_R^h + \mu_L^e} \right),
\end{gather*}
which determines the location of the onset of finite conductance corresponding to the yellow point in Fig.~\ref{fig_2}\textbf{b,c,d}.
When increasing $V_\text{isl}$ (and thus $N_{\text{g}}$) but keeping $e|V|=2\Delta$, the next point with finite conductance is expected at the blue point in Fig.~\ref{fig_2}\textbf{b,c,d}, where it has to be noted that the order of the tunneling processes is reversed, and the corresponding conditions read in this case
\begin{gather*}
    \mu_L^e + \mu_R^h = 2\Delta,\\
    N_{\rm g} = \frac{1}{2} \left( 3 + 2N_c + \frac{\Delta}{E_\text{c}} \frac{\mu_R^h - \mu_L^e}{\mu_R^h + \mu_L^e} \right).
\end{gather*}
We thus see that these points of lowest-bias finite conductance always occur at $e|V| = 2\Delta$ and are regularly spaced by $1e$, independent of the bias asymmetry, i.e., the ratio $\mu_L^e / \mu_R^h$.
The exact position of the points does, however, depend on the bias asymmetry.

The numerical results of a full transport calculation in the weak coupling regime are presented in Figs.~\ref{fig_2}\textbf{d,e} for the 0L and 1L, using the model and ingredients introduced above. We see that the model reproduces the main features observed in the experiment, see the top row in Fig.~\ref{fig_1}. For these calculations we used $E_{\rm c} = 3\Delta_0$.

The situation is different in the strong coupling regime, where our measurements consistently show signatures of subgap states, see Figs.~\ref{fig_0}\textbf{f}--\textbf{i} and the bottom row of Fig.~\ref{fig_1}, in particular Fig.~\ref{fig_1}\textbf{j}. We take this as an indication that in this regime, the tunneling rates into possible subgap states have become comparable to the total tunneling rates into the continuum of states.

To describe the strong coupling situation, we thus include the effect of discrete subgap states into our description. For simplicity, we add only one state with energy $\varepsilon < \Delta$ and assume that it has significant (equal) tunnel coupling to the source and drain leads, allowing for efficient tunneling in and out of quasiparticles. We note that higher-order processes are not included in our rate-equation approach, so it does not describe cotunneling features or the charging energy renormalization, which we effectively include by reducing $E_\text{c}$.
When $\varepsilon<E_\text{c}$, the inclusion of this state can open up a transport channel at the even--odd charge degeneracy points, leading to zero-bias conductance features.
The spacing and periodicity of the zero-bias features depend on the magnitude of $\varepsilon$ compared to $E_\text{c}$~\cite{albrecht_exponential_2016,Albrecht_PRL2017}.

In the 0L, see Fig.~\ref{fig_1}\textbf{j}, we observe zero-bias peaks that are approximately $2e$-spaced. The NDC appearing at small bias signals the presence of nearby excited odd states that can block the current.
In our model, the continuum states could play this role since the fast intra-continuum relaxation causes the tunneling rates into and out of the continuum to be vastly different, also in the strong coupling limit.
Thus, as soon as incoming electrons have enough energy to excite quasiparticles in the continuum, the electron flow is quenched until the leads can provide enough energy to excite two quasiparticles above the gap, that can recombine into a Cooper pair, leading to a region of NDC.
The relatively low $V$ at the onset of the NDC indicates that $\Delta$ is most likely only marginally larger than $\varepsilon$.
Similar signatures have been reported before as arising from bound states with energies close to $\Delta$~\cite{Higginbotham_NatPhys2015,Albrecht_PRL2017}.

In the 1L, see Fig.~\ref{fig_1}\textbf{l}, the picture is qualitatively different.
Here, the sample also shows low-energy transport, but (i) no NDC features can be observed at low bias and (ii) the periodicity of the zero-bias features seems to be doubled compared to the one in the 0L.
Both these observations indicate the presence of low-energy states that contribute to transport~\cite{valentini2022majorana}. 

Figures~\ref{fig_2}\textbf{f,g} show the calculated conductance using the same model as in Figs.~\ref{fig_2}\textbf{d,e}, but adding a significant tunneling rate to the subgap state. In the 0L, Fig.~\ref{fig_2}\textbf{f}, the subgap state is considered to be close to the gap ($\varepsilon=0.95\Delta$) and in the 1L, Fig.~\ref{fig_2}\textbf{g}, we set it at zero energy. In both cases, we assumed that $E_\text{c}= 1.25\, \Delta$.
We see that our model indeed captures most qualitative features in the data discussed above. Our simulations based on a few simple ingredients reproduce all qualitative features of the data and thus provide a consistent understanding of the behavior of the device

\section{Conclusion}

In this study, we performed tunnelling and finite bias spectroscopy of an InAs/Al full-shell island for different tunnel barrier configurations to explore the subgap physics of the hybrid superconductor--semiconductor island.
The experimental data show that the presence or absence of subgap features depends strongly on the coupling strength to the leads.
With local tunneling spectroscopy, we observed signatures of subgap states only in the 1L and when in the strong coupling regime. Coulomb spectroscopy at strong coupling not only confirmed the presence of extended subgap states in the 1L, but also suggested the existence of a subgap state with energy close to the superconducting gap in the 0L. The subgap states' dependence on the applied flux and their interaction with tunnel barriers suggest compatibility with vortex-like states, whose visibility might be influenced by the tunnel barrier, as theoretically proposed in Ref.~\cite{san2023theory}.
Our work emphasizes the importance of tunable tunnel barriers, in order to understand the complex phase space of subgap states in hybrid devices.

\section{acknowledgements}
This research was supported by the Scientific Service Units of ISTA, through resources provided by the MIBA Machine Shop and the Nanofabrication facility. This research and related results were made possible with the support of the FWF Project with DOI:10.55776/F86. We acknowledge support from the European Research Council under the European Unions Horizon 2020 research and innovation programme under Grant Agreement No. 856526, the Swedish Research Council under Grant Agreement No. 2020-03412, the Spanish Comunidad de Madrid (CM) ``Talento Program'' (Project No. 2022-T1/IND-24070), the Spanish Ministry of Science, innovation, and Universities through Grant PID2022-140552NA-I00 and NanoLund.

\clearpage
\appendix
\renewcommand{\thefigure}{S\arabic{figure}}
\setcounter{figure}{0}

\onecolumngrid

\section{Additional data}
\label{app:data}

\begin{figure*}[b]
    \includegraphics[]{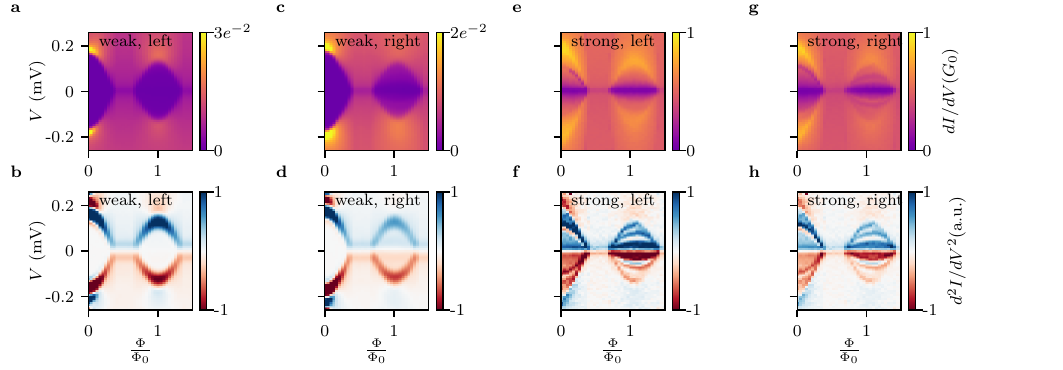}
    \caption{\textbf{Tunneling spectroscopy data for device B.} \textbf{a} [\textbf{b}]  $dI/dV$ [$d^2I/dV^2$] of device A as a function of voltage bias $V$ and flux $\Phi$ with the left side tuned to weak coupling ($G_{\textrm{L}} \approx 0.01\, G_0$) and the right side tuned to be fully transparent (resulting in $E_{\textrm{c}} \approx 0$).  \textbf{c} [\textbf{d}] $dI/dV$ [$d^2I/dV^2$] as a function of $V$ and $\Phi$ with the right side tuned to weak coupling ($G_{\textrm{R}} \approx  0.008\, G_0$) and the left side fully transparent.  \textbf{e} [\textbf{f}] $dI/dV$ [$d^2I/dV^2$] as a function of $V$ and $\Phi$ with the left side tuned to strong coupling ($G_{\textrm{L}} \approx  0.6\, G_0$) and the right side fully transparent. \textbf{g} [\textbf{h}] $dI/dV$ [$d^2I/dV^2$] as a function of $V$ and $\Phi$ with the right side tuned to strong coupling ($G_{\textrm{R}} \approx  0.55\, G_0$) and the left side fully transparent. A Savitzky-Golay filter was used to improve the readability of the $d^2I/dV^2$ maps.}
    \label{figuresup_tun}
\end{figure*}

\begin{figure*}[b]
    \vspace{4em}
    \includegraphics[]{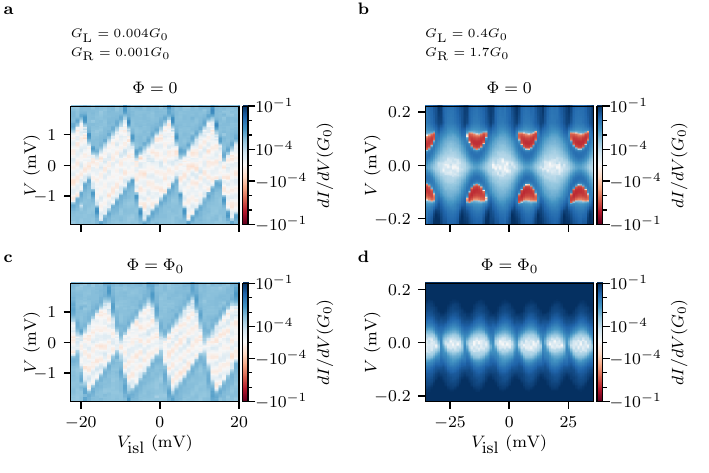}
    \caption{\textbf{Coulomb spectroscopy data for device A.} First [second] row corresponds to $\Phi=0$ [$\Phi=\Phi_0$].}
    \label{figuresup_coul}
\end{figure*}

\begin{figure*}
    \includegraphics[]{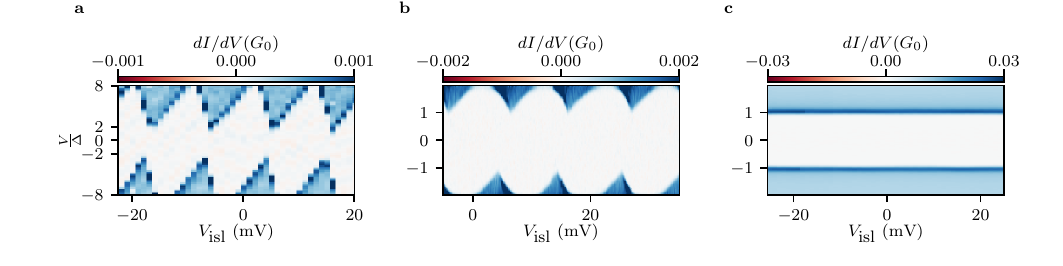}
    \caption{\textbf{Crossover from Coulomb to tunneling spectroscopy at $\Phi=0$.} Coulomb diamonds for different coupling strengths. In \textbf{a} both tunnel barriers are in the weak coupling regime ($G_{\textrm{L}} = 0.001\, G_0$ and $G_{\textrm{R}} = 0.004\, G_0$). Thus, transport is allowed only if $e| V | > 2 \Delta$. In \textbf{b}, the junctions are asymmetrically tuned ($G_{\textrm{L}} = 1.6\, G_0$ and $G_{\textrm{R}} = 0.006\, G_0$) such that transport is allowed only if $e| V | > \Delta$, resembling a tunneling spectroscopy experiment. In \textbf{c} one junction is in the weak coupling regime ($G_{\textrm{L}} = 0.01\, G_0$ and the other one is open $G_{\textrm{R}} \gg 2\, G_0$), leading to negligible effects of $E_{\textrm{c}}$. Data in \textbf{a} and \textbf{b} are taken from the $400$ nm long island and data of \textbf{c} are taken from a $900$ nm long island.}
    \label{crossover}
\end{figure*}

\begin{figure*}
    \includegraphics[]{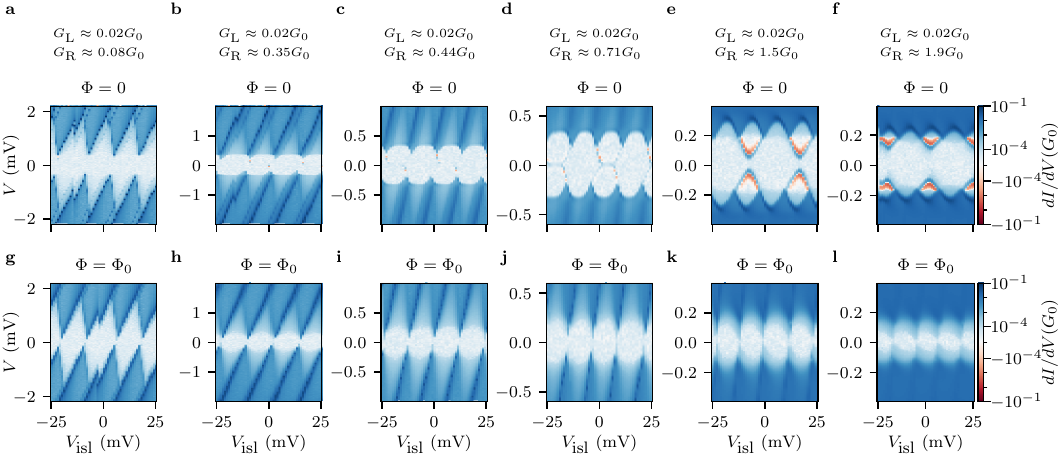}
    \caption{\textbf{Diamond evolution for asymmetric tunnel barriers (device B).} First [second] row corresponds to $\Phi=0$ [$\Phi=\Phi_0$].
    We note that the skewness of the Coulomb diamonds initially decreases with increasing $G_{\rm R}$ and then seems to increase again, in the opposite direction.
    A more systematic study of the shape of the Coulomb diamonds as a function of $G_{\rm L,R}$ and the bias direction could reveal the detailed dependence of the island's capacitance matrix on all gate voltages.}
    \label{sup2}
\end{figure*}

\begin{figure*}
    \includegraphics[]{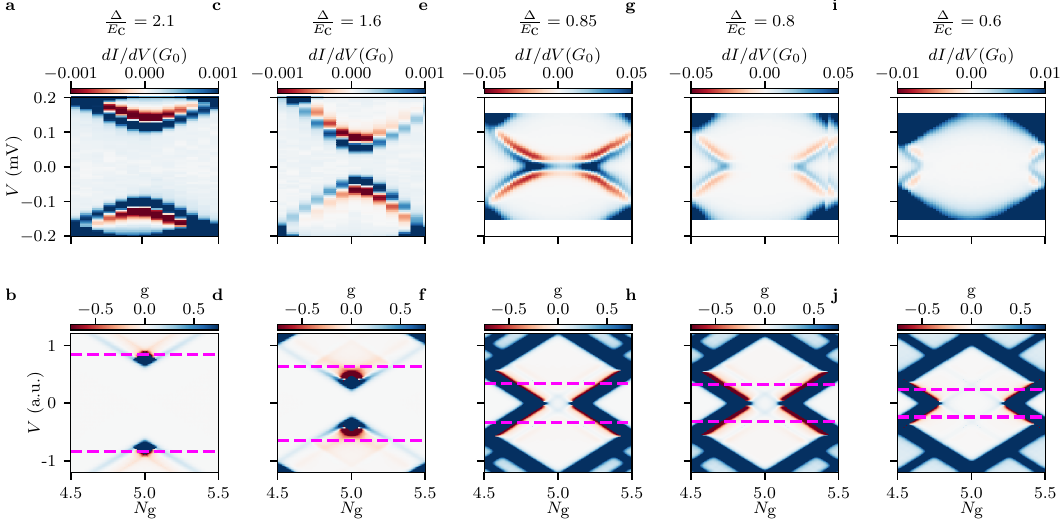}
    \caption{\textbf{Interplay between $E_{\textrm{c}}$ and $\Delta$ in the strong coupling regime (device B).} \textbf{Top row} $dI/dV$ as a function of  $V$ and of $N_{\textrm{g}} \propto V_{\textrm{isl}}$ in the strong coupling regime for different value of $E_{\textrm{c}}$ and $\Delta$. In \textbf{a}, $G_{\textrm{L}} \approx 0.01 \,G_0$, $G_{\textrm{R}} = 1.6 \,G_0$, $\Phi=0$, $\Delta = 180 \ \mu\textrm{eV}$. In \textbf{c}, $G_{\textrm{L}} \approx 0.01\, G_0$, $G_{\textrm{R}} \approx 0.9\, G_0$, $\Phi=0$, $\Delta = 180 \ \mu\textrm{eV}$. In \textbf{e}, $G_{\textrm{L}} \approx 0.6\, G_0$, $G_{\textrm{R}} \approx 0.55\, G_0$, $\Phi=0.125\, \Phi_0$, $\Delta = 150 \ \mu\textrm{eV}$. In \textbf{g}, $G_{\textrm{L}} \approx 0.6\, G_0$, $G_{\textrm{R}} \approx 0.55\, G_0$, $\Phi=0.17\, \Phi_0$, $\Delta = 130 \ \mu\textrm{eV}$. In \textbf{i}, $G_{\textrm{L}} \approx 0.6\, G_0$, $G_{\textrm{R}} \approx 0.55\, G_0$, $\Phi=0.25\, \Phi_0$, $\Delta = 115 \ \mu\textrm{eV}$. \textbf{Bottom row} Calculated $g$ as a function of  $V$ and $N_{\text{g}}$ for the same parameters, assuming that $\Delta-\varepsilon=0.05\,\Delta_0$. The dashed magenta horizontal lines indicate the values of $\pm \Delta$.}
    \label{fig_3}
\end{figure*}


\twocolumngrid
\clearpage

\section{Interplay between $E_{\textrm{c}}$ and $\Delta$}
\label{app:interplay}

In order to understand how the even--odd spacing of Coulomb peaks emerges, we study the interplay between $E_{\textrm{c}}$ and $\Delta$ in the strong coupling regime, where transport is dominated by subgap states.
We focus on the 0L, where the energy of the subgap state seems to be close to $\Delta$.
 
To ensure that at least one junction is in the strong coupling regime and to tune the device from the superconductivity-dominated regime ($\Delta > E_{\textrm{c}}$) to the Coulomb-dominated regime ($\Delta < E_{\textrm{c}}$), we adjust both junctions and, additionally, $\Phi$.

First, in Fig.~\ref{fig_3}\textbf{a}, we tune the couplings asymmetrically ($G_{\textrm{L}} \approx 0.01\, G_0$ and $G_{\textrm{R}} \approx 1.6\, G_0$) such that $E_\textrm{c} \approx 80~\mu$eV, i.e., $\Delta/E_{\textrm{c}}\approx2.1$ at $\Phi=0$.
Transport at $e| V| < \Delta$ is still observed, indicating the presence of a subgap bound state, similarly to Fig.\ref{fig_1}\textbf{j}.
However, the typical arc shape symmetric in $V=0$ accompanied by NDC is pushed to higher energies, suggesting that both $\Delta$ and $\varepsilon$ are indeed larger than $E_{\rm c}$. 
Next, we increase the charging effects slightly by decreasing the junction transparencies, such that $\Delta/E_{\textrm{c}}\approx 1.6$ (Fig.~\ref{fig_3}\textbf{c}). Now, the transport arcs move to lower values of $V$. This is due to a reduction of the excitation energy involving quasiparticles excited in the subgap state. Eventually, the arcs will merge when $\varepsilon=E_{\textrm{c}}$, yielding finite zero-bias transport at the charge-degeneracy points $N_\textrm{g} = 2n+1$.

Next, we further reduce $\Delta/E_{\textrm{c}}$ below 1; to this end, we tune the junctions symmetrically ($G_{\textrm{L}} \approx 0.6\, G_0$ and $G_{\textrm{R}} \approx 0.55\, G_0$) such that $E_\textrm{c} = 180\ \mu$eV and then we decrease the superconducting gap by applying a small magnetic field $\Phi=0.125\,\Phi_0$, altogether resulting in $\Delta/E_{\textrm{c}}\approx 0.85$. In this case, the arcs merge at zero bias, symmetrically with respect to $N_\textrm{g} = 2n+1$, giving an even--odd spacing of the Coulomb peaks~\cite{lafarge1993measurement,Higginbotham_NatPhys2015}. The Coulomb peaks move further apart when the ratio $\Delta/E_c$ is decreased further, which we probe by increasing $\Phi$ further, see Figs.~\ref{fig_3}\textbf{g,i}. All these observations are reproduced qualitatively by our theoretical model, see Figs.~\ref{fig_3}\textbf{f,h,j}, where we used the same ratios of $\Delta/E_{\rm c}$ as extracted from the experiments.

\section{Theoretical model}\label{sec:model}

We describe the superconducting island coupled to superconducting leads using the Hamiltonian
\begin{equation}
    H=H_{\rm I}+H_{\rm L}+H_{\rm T}\,,
\end{equation}
where the leads are described by 
\begin{equation}
    H_{\rm L}=\sum_{\rm \nu,k,\sigma}\left( \xi_{\nu k\sigma}-\mu_\nu \right)c^{\dagger}_{\nu k\sigma}c_{\nu k\sigma}\,.
\end{equation}
Here, $\xi_{\nu k\sigma}$ represents the electron energy, and $c_{\nu k\sigma}$ is the annihilation operator for an electron in lead $\nu=L,\,R$ with momentum $k$ and spin $\sigma \in {\uparrow, \downarrow}$. We assume that each lead is in internal equilibrium, described by a Fermi-Dirac distribution $n_{\rm{F}}$ with chemical potential $\mu_\nu$ and temperature $T$.

The superconducting island is described by
\begin{equation}
    H_{\rm I}=\sum_{j,\sigma}\varepsilon_{j\sigma} \gamma_{j\sigma}^\dagger \gamma_{j\sigma}+E_{\rm el}(N)\,.
    \label{H_island}
\end{equation}
In this equation, $j$ indexes a state with single-particle energy $\varepsilon_{j\sigma}$. In this work, we consider two types of states. We first consider a discrete state with energy below the superconducting gap ($j=0$), $\varepsilon_{0\sigma} < \Delta$. We also include the continuum of states on the island ($j>0$), appearing for energies above the superconducting gap $\Delta$ due to proximity to the superconductor. We describe the continuum states using the conventional BCS density of states
\begin{equation}
    \varepsilon_{j\sigma}(\omega_{j})=\sqrt{\Delta^2+\omega_{j}^2}\,,
\end{equation}
where $\omega_j$ is the energy of $j$-th quasiparticle state. The Bogoliubov--de Gennes (BdG) operators are given by
\begin{equation}
    \gamma_{j\sigma}=u_j d_{j\sigma}-\rho_\sigma \, v_je^{-i\phi}d_{j\bar{\sigma}}^\dagger\,,
\end{equation}
where, $d$ and $d^\dag$ represent the annihilation and creation operators of electrons on the island, $e^{-i\phi}$ annihilates a Cooper pair on the island (where $\phi$ is the phase operator of the induced superconducting order parameter in the semiconductor), and $\rho_\sigma=\pm$ for $\sigma=\uparrow,\downarrow$. Here, $u_j$ and $v_j$ are the BdG coefficients of the states. For the continuum of states, we use the conventional BCS expressions, given by $u_j^2=(1+\omega_j/\varepsilon_{j})/2$ and $u_j^2+v_j^2=1$. For the BdG coefficients of the subgap state, we take $u_0^2=v_0^2=1/2$, although the described qualitative features do not rely on this specific choice.

Due to electron confinement, the island shows a charging energy given by 
\begin{equation}
    E_{\rm el}(N)=E_{\rm c}(N-N_{\rm g})^2\,,
\end{equation}
Here, $N_{\rm g}$ represents the dimensionless gate-induced charge offset, and $N$ is the (excess) electron number operator, taking into account the fermionic occupation of the states and the number of Cooper pairs on the island, $N=N_c+n_s+n_c$.

Finally, the tunneling between the leads and the superconducting island  is described by the Hamiltonian
\begin{equation}
    \begin{split}
    H_T & =\sum_{\nu, k,j,\sigma} \left(t_{\nu kj\sigma}c_{\nu k\sigma}^\dagger d_{j\sigma}+ {\rm H.c.}\right).
	\end{split}
\end{equation}
Here, $t$ represents the tunneling amplitude. We define the corresponding tunneling rates as $\Gamma_{\nu j\sigma}=2\pi\rho_{F\nu} |t_{\nu j\sigma}|^2$, assuming them to be $k$-independent in the wideband limit, where $\rho_{F\nu}$ is the density of states at the Fermi level of lead $\nu$.

\subsection{Formalism}\label{ssec:formalism}
In this work, we consider the limit where the tunneling rates are the smallest energy scales, so we can use a perturbative treatment using first-order rate equations, $\Gamma_{\nu j\sigma}\ll T$, for the details see for example Ref.~\cite{Souto_PRB2022}.

For our transport calculations, we label the relevant state of the island as
\begin{equation}
    \left|a\right\rangle=(N_c,n_s,n_c),
\end{equation}
where $N_c/2$ is the number of Cooper pairs in the island and $n_s$ and $n_c$ describe the occupation of the subgap and continuum excited states. In this work, we 
assume that coherences between different states in the superconducting island are negligible. Therefore, the system is fully characterized by occupation probabilities $P_a$, and their time evolution is given by the master equation
\begin{equation}\label{eq:master}
	\dot{P}_a=\sum_b\left[-\Gamma_{a\to b}P_a+\Gamma_{b\to a}P_b\right]\,.
\end{equation}
In the stationary limit, the system does not evolve in time and we can impose $\dot{P}^{\rm{stat}}_a=0$. These conditions, together with normalization  $\sum_b P^{\rm{stat}}_a=1$, form a complete linear system of equations for the stationary occupation probabilities of the island states. Using the resulting $P^{\rm{stat}}_a$, the current flowing from lead $\nu$ to the device is determined by the island transition rate to other states as
\begin{eqnarray}\label{eq:current}
        I_\nu=\sum_{a,b}s_{a\to b}\,\Gamma_{\nu,a\to b}
	     P^{\rm{stat}}_a\,,
\end{eqnarray}
where $\Gamma_{\nu,a\to b}$ is the transition rate between $\left|a\right\rangle$ and $\left|b\right\rangle$ involving an electron tunneling to or from lead $\nu=L,R$, and the sum runs over all the possible tunneling processes. Here, we take $s=+1$ ($s=-1$) in case an electron tunnels in (out) of the island from (to) the lead $\nu$ when transitioning between from $\left|a\right\rangle$ to $\left|b\right\rangle$.

In Eqs.~\eqref{eq:master} and \eqref{eq:current} (and using the approximation ${\cal T} = H_{\rm T}$) we are only considering processes that change the island occupation by $\pm$1 electron. Therefore, we disregard cotunneling features and processes changing the island occupation by more than 1 electron, like Andreev reflection. These two processes are important to quantitatively describe the open regime, when the $\Gamma$'s become comparable to other energy scales.

\begin{widetext}
\subsection{Tunneling rates}

The tunneling rates between the leads and the island can then be calculated. For the tunneling between the leads and the discrete subgap state, they are given by
\begin{align}
    \begin{split}
         	\Gamma^{\rm s}_{\nu,\left|N_c,n_s,n_c\right\rangle\to\left|N_c,n_s+1,n_c\right\rangle}&=\gamma_{\nu,0} \left|u_0\right|^2 n_{\rm F}(E_f-E_i-\mu_\nu),\\
	        \Gamma^{\rm s}_{\nu,\left|N_c,n_s,n_c\right\rangle\to\left|N_c,n_s-1,n_c\right\rangle}&=\gamma_{\nu,0} \left|u_0\right|^2 n_{\rm F}(\mu_\nu+E_f-E_i),\\
	        \Gamma^{\rm s}_{\nu,\left|N_c,n_s,n_c\right\rangle\to\left|N_c+2,n_s-1,n_c\right\rangle}&=\gamma_{\nu,0} \left|v_0\right|^2 n_{\rm F}(E_f-E_i-\mu_\nu),\\
         	\Gamma^{\rm s}_{\nu,\left|N_c,n_s,n_c\right\rangle\to\left|N_c-2,n_s+1,n_c\right\rangle}&=\gamma_{\nu,0} \left|v_0\right|^2 n_{\rm F}(\mu_\nu+E_f-E_i),
    \end{split}
    \label{SeqRates_ABS}
\end{align}
where $E_i$ and $E_f$ are the initial and final energies of the island and $n_F$ is the Fermi distribution function. The two first rates in Eq.~\eqref{SeqRates_ABS}, proportional to $|u_0|^2$, describe the normal tunneling in/out of electrons between the lead and the island. The second two terms, proportional to $|v_0|^2$, represent processes where additionally a Cooper pair splits/recombines after the tunneling of an electron.

Electrons have many available states to tunnel into in the continuum of states of the island above the gap. To account for this, we have to integrate over the above-gap density of states. The total rate for one electron tunneling in to the continuum is thus given by~\cite{Albrecht_PRL2017}
\begin{equation}
    \Gamma^{\rm c}_{\nu,\left|N_c,n_s,n_c\right\rangle\to\left|N_c,n_s,n_c+1\right\rangle}=\gamma^{\rm in}_{\nu,c}\int_{0}^{\infty} d\omega A(\omega)n_{\rm F}(E_f-E_i-\mu_\nu)\,,
\end{equation}
where the density of states of the superconductor is given by
\begin{equation}
    A(\omega)={\rm Im}\left[\frac{\omega}{\sqrt{\Delta^2-(\omega+i\eta)^2}}\right]\,,
\end{equation}
where $\eta$ is the Dynes parameter broadening the density of states, set to $0.002\,\Delta_0$, with $\Delta_0$ being the gap at zero applied magnetic field.
Breaking of a Cooper pair is another process that can excite a quasiparticle above the continuum. This process is described by a similar tunnel rate, given by
\begin{equation}
    \Gamma^{\rm c}_{\nu,\left|N_c,n_s,n_c\right\rangle\to\left|N_c-2,n_s,n_c+1\right\rangle}=\gamma^{\rm in}_{\nu,c}\int_{\Delta}^{\infty} d\omega A(\omega)n_{\rm F}(\mu_\nu+E_f-E_i)\,.
\end{equation}
In these expressions we have assumed a very low density of quasiparticles in the superconducting island, in such a way that the available states in the continuum do not fluctuate.

The rates for tunneling out are given by
\begin{align}
    \begin{split}
	        \Gamma^{\rm c}_{\nu,\left|N_c,n_s,n_c\right\rangle\to\left|N_c,n_s,n_c-1\right\rangle}&=\gamma^{\rm out}_{\nu,c} \left|u_c\right|^2 n_{\rm F}(\mu_\nu+E_f-E_i),\\
	        \Gamma^{\rm c}_{\nu,\left|N_c,n_s,n_c\right\rangle\to\left|N_c+2,n_s,n_c-1\right\rangle}&=\gamma^{\rm out}_{\nu,c} \left|v_c\right|^2 n_{\rm F}(E_f-E_i-\mu_\nu).
    \end{split}
\end{align}
Similarly to the case of the subgap state we choose $|u_c|^2=|v_c|^2 = \frac{1}{2}$, consistent with the BCS prediction for the edge of the continuum. Again, our qualitative results do not depend on this choice.

Finally, we consider processes where two quasiparticles in the continuum of states can recombine forming a Cooper pair. This process, that does not conserve energy, is essential to understand some of the features, including the open Coulomb diamonds in the regime where $E_\text{c}>\Delta$. The rate for Cooper pair recombination is included as
\begin{equation}
    \Gamma^{\rm r}_{\left|N_c,n_s,n_c\right\rangle\to\left|N_c+2,n_s,n_c-2\right\rangle}=\gamma^{\rm r}\,.
\end{equation}
This process conserves the charge on the island and does not contribute directly to the current in Eq.~\eqref{eq:current}. 
\end{widetext}


\bibliography{bibliography}

\end{document}